# SPACECRAFT CHARGING: INCOMING AND OUTGOING ELECTRONS

Shu T. Lai, SPL, MIT, Cambridge, 02139, USA

*Abstract*

This paper presents an overview of the roles played by incoming and outgoing electrons in spacecraft surface and stresses the importance of surface conditions for spacecraft charging. The balance between the incoming electron current from the ambient plasma and the outgoing currents of secondary electrons, backscattered electrons, and photoelectrons from the surfaces determines the surface potential. Since surface conditions significantly affect the outgoing currents, the critical temperature and the surface potential are also significantly affected. As a corollary, high level differential charging of adjacent surfaces with very different surface conditions is a space hazard.

## INTRODUCTION

The most important region for spacecraft charging is the geosynchronous region, where the electrons are often of high energy (keV) depending on the space weather and many satellites are there. Measurements in that region have shown that the flux of electrons is nearly two orders of magnitude higher than that of ions. Geosynchronous satellites often charge to negative voltages during adverse space weather.

## INCOMING ELECTRONS

In plasmas, electrons are much faster than ions because of their mass difference. This is true in space and in the laboratory. If one puts an initially unchanged spacecraft in space, the spacecraft will likely intercept more incoming electrons than incoming ions. As a result of intercepting more electrons, the spacecraft charges to a negative potential. The level of spacecraft charging at equilibrium is determined by current balance. That is, the sum of all currents to the spacecraft equals zero.

## OUTGOING ELECTRONS

For every incoming primary electron of energy E, there are $\delta(E)$ outgoing secondary and $\eta(E)$ backscattered electrons. The probabilities, $\delta(E)$ and $\eta(E)$, are called secondary electron yield (SEY) and backscattered electron yield (BEY) respectively. They are also called secondary electron emission coefficient and backscattered electron emission coefficient respectively. Their properties are known to depend not only on E but also on the surface material [1,2,3,4]. Graphs of $\delta(E)$ and $\eta(E)$ for typical spacecraft surface materials are shown in Fig.1.

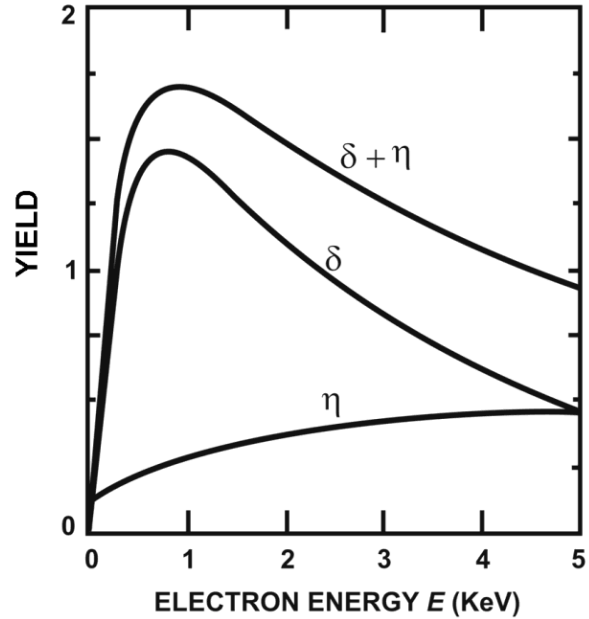

Figure 1. Yields of secondary and backscattered electrons induced by the impact of primary electrons of energy *E*.

In Fig.1, the $\delta(E)$ graph starts at 0 at E=0, rises to the maximum $\delta_{max}(E)$ at $E=E_{max}$, and decreases monotonically as E increases. For most materials, $\delta_{max}(E)$ exceeds unity and the graph $\delta(E)$ has two unity crossings at $E=E_1$ and $E=E_2$. Typically, $E_1$ is about 40eV and $E_2$ about 1600eV. Beyond $E_2$, $\delta(E)$ is less than unity. The $\eta(E)$ curve is always below unity. Secondary electrons are much more abundant than backscattered electrons. Together, the sum of $\delta(E)$ and $\eta(E)$ contribute to the outgoing electron current. If there are other currents, such as photoemission from surfaces in sunlight or artificial charged particle beam emissions, the currents have to be included in the current balance.

## MAXWELLIAN SPACE PLASMA

Plotting the log of a Maxwellian electron distribution *f*(E) as a function of the primary electron energy E, one obtains a straight line whose slope is -1/kT, where k is Boltzmann's constant and T the electron temperature (Fig.2). In Fig.2, one can consider two camps of electrons coming to the spacecraft surface. The low-energy camp favors positive voltage charging, whereas the high-energy camp favors negative charging.

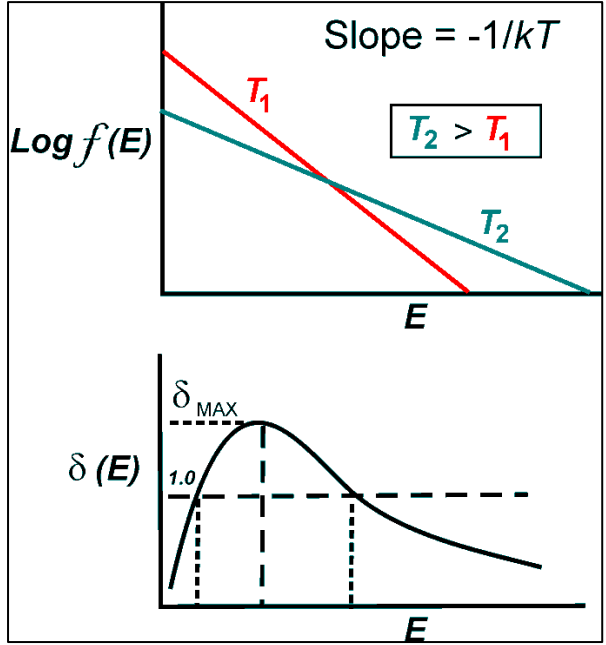

Figure 2. (Upper) Slope of log *f(E)*. There are more hot electrons in a high temperature distribution *f(E)*. (Lower) Hot electrons are responsible for negative voltage charging negative voltages occurs.

## CRITICAL TEMPERATURE

At low temperatures, there are more low-energy electrons than high-energy electrons. Suppose the temperature is initially low and the spacecraft is uncharged. Since secondary electrons are of a few eV only, positive charging by secondary electron emission is up to a few volts only. Since charging to a few volts is harmless, we can ignore it and regard it as practically uncharged. Now, suppose the temperature T is increasing, the slope (-1/kT) decreases accordingly (Fig.2), and therefore there are more and more hot electrons. Eventually, there must exist a critical temperature, T=T*. Above T*, charging to negative voltages occurs; below T*, charging to negative voltages does not occur. As temperature increases above T*, the charging level increases. Indeed, charging to -keV occurs at geosynchronous altitudes during severe space weather [5,6].

## ONSET OF SPACECRAFT CHARGING

To study the onset of charging, we ignore the ambient ion current because it is two orders of magnitude smaller than the ambient electron current. Let us also ignore photoelectrons. In this simple model, the players are incoming and outgoing electrons only. For normal incidence, the current balance equation [Appendix] is given as follows.

$$\int_0^\infty dE E f(E) = \int_0^\infty dE E f(E)[\delta(E) + \eta(E)] \quad (1)$$

where the Maxwellian distribution function $f(E)$ is given by

$$f(E) = n(m/2\pi kT)^{3/2} \exp(-E/kT) \quad (2)$$

Substituting eq(2) into eq(1), one finds that the electron density *n* cancels out on both sides, because for more electrons coming in, there are more secondary and backscattered electrons going out. We have therefore two simple, but useful, properties in this model. They are (1) the onset of charging is independent of the electron density, and (2) for a given surface material, the solution of eq(1) is T = T*, the critical electron temperature for the onset of spacecraft charging. In simple words, whenever the electron temperature in a Maxwellian plasma in space exceeds the critical temperature, negative voltage spacecraft charging occurs and the occurrence is independent of the electron density.

Note that if the incoming electrons are at various incidence angles, one needs to include the integration over angles in eq(1). If other currents such as photoelectrons and beam electrons are involved, they have to be included. If there is blockage of currents, it has to be taken into account. If the space plasma deviates widely from being Maxwellian, temperature is undefined and one needs to use other parameters.

To calculate the numerical value of T*, one needs to know the functions, δ(E) and η(E). There are many δ(E) and η(E) functions published in the literature. Fig.4 shows a comparison of results using various functions. If we know which ones are the best, we would use them in eq(1).

In recent years, two advances made by CERN electron cloud researchers have impacted the current balance studies in spacecraft charging. (1) Furman [7] proposed a δ(E) formula that depends on the surface condition parameterized by *s*.

$$\delta(E) = \delta_{max} \frac{s(E/E_{max})}{s - 1 + (E/E_{max})^s} \quad (3)$$

$$E_{max}(\theta) = E_{max}(0)[1 + 0.7(1 - \cos\theta)] \quad (4)$$

$$\delta_{max}(\theta) = \delta_{max}(0) \exp[0.5(1 - \cos\theta)] \quad (5)$$

where θ is the angle of incidence. The surface parameter *s* significantly affects the value of δ(E) and therefore spacecraft charging (Fig.3). (2) The η(E) function rises to unity as E approaches 0 [8,9]. This new property does not have much effect on negative voltage charging because the primary electron energy involved is near zero. It can influence positive charging for some materials at low primary electron energies.

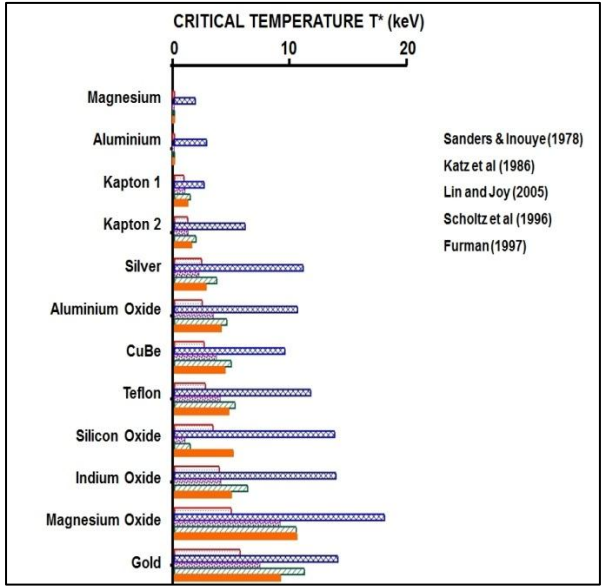

Figure 3. Critical temperature for the onset of spacecraft charging computed by using various δ(E) functions.

## CHARGING OF MIRRORS

Since the surface condition significantly affects the secondary electron yield and backscattered electron yield, they, in turn, affect spacecraft charging. As an interesting example, a highly reflecting mirror in sunlight should emit no photoelectrons, because there is too little photon energy imparted to the mirror for photoemission. As a corollary, we conjecture that a mirror should charge to negative potentials in sunlight as if it were in eclipse. It is worthwhile to do laboratory experiments for proving or refuting this idea of charging of mirrors in space.

As an example, if a solar panel is flanked by mirrors for focusing more sunlight onto the panel, differential charging between the mirror and the panel may occur because of the vastly different surface conditions [10].

On the other hand, the outgoing electrons from a very rough surface is also reduced. Photoelectrons are of low energy (a few eV). Suppose they are emitted from the deep and rough valleys of the surface. They impact on the valley walls but are not energetic enough to generate secondary electrons.

## CONCLUSION

Low energy electrons emitted from surfaces are important in various fields. This paper gives an overview of spacecraft surface charging, which is controlled by current balance between the incoming and outgoing currents. Secondary and backscattered electron currents calculated by using the yield functions obtained from handbooks or journals is inadequate. Surface conditions are very important. One needs to measure the surface condition for accurate calculations.

## ACKNOWLEGMENT

The author gratefully thanks Frank Zimmermann of CERN and Roberto Cimino of LNF for inviting this paper and Manuel Martinez-Sanchez of MIT for hospitality and discussions.

## APPENDIX

The electron flux $J$ arriving at a surface is given as follows.

$$J = qn\text{v} \quad (A.1)$$

where $n$ is the electron density, $q$ the electron charge, and v the electron velocity. If the electron velocity distribution is $f(\text{v})$, the electron flux $J$ is given as follows.

$$J = q\int_0^\infty d^3\text{v} f(\text{v})\,\text{v} \quad (A.2)$$

where v is the electron velocity. In polar coordinates, the flux $J$ of eq(A.2) is written as follows.

$$J = q\int_0^\infty d\text{v}\,\text{v}^2 \int_0^{\pi/2} d\varphi \int_0^{2\pi} d\theta \sin\varphi f(\text{v})\text{v} \quad (A.3)$$

The Maxwellian velocity distribution, $f(\text{v})$, is of the following form:

$$f(\text{v}) = n\left(\frac{m}{2\pi kT}\right)^{3/2} \exp\left(-\frac{m\text{v}^2}{2kT}\right) \quad (A.4)$$

where $m$ is the electron mass, $k$ the Boltzmann constant, and $T$ the electron temperature. Since electrons are measured as a function of energy $E$, it is convenient to use $E$ as the variable instead of v. Let us denote $f(E)$ as the electron velocity distribution where $E = (1/2)m\text{v}^2$.

$$f(E) = n\left(\frac{m}{2\pi kT}\right)^{3/2} \exp\left(-\frac{E}{kT}\right) \quad (A.5)$$

Using $E$, the incoming electron flux $J$ in eq(A.3) is written in the following form:

$$J = q\int_0^\infty dE\,E \int_0^{\pi/2} d\varphi \int_0^{2\pi} d\theta \sin\varphi f(E) \quad (A.6)$$

For normal incidence, we need not elaborate the angular dependence of the secondary and backscattered electron yields. The balance between the outgoing and incoming electron fluxes can be written as follows.

$$q\int_0^\infty dE\,E \int_0^{\pi/2} d\varphi \int_0^{2\pi} d\theta \sin\varphi f(E)$$
$$= q\int_0^\infty dE\,E \int_0^{\pi/2} d\varphi \int_0^{2\pi} d\theta \sin\varphi f(E)\left[\delta(E) + \eta(E)\right]$$
$$(A.7)$$

Since *q* and the angles in eq(A.7) cancel out on both sides, the electron flux balance equation becomes

$$\int_0^\infty dE\, E f(E) = \int_0^\infty dE\, E f(E)[\delta(E) + \eta(E)] \quad (A.8)$$

which is eq(1) on page 2.